\documentclass[twocolumn,aps,pra,superscriptaddress,longbibliography]{revtex4-2}

\usepackage[utf8]{inputenc}
\usepackage{natbib}
\usepackage[T1]{fontenc}
\usepackage{babel}
\usepackage{float}
\usepackage{amsmath}
\usepackage{amssymb}
\usepackage{amsfonts}
\usepackage{dsfont}
\usepackage{xcolor}
\usepackage{bbm}
\usepackage[caption=false,font=normalsize,labelfont=sf,textfont=sf]{subfig} 
\usepackage{graphicx,epstopdf}
\usepackage{url}
\usepackage{textcase}
\usepackage{bm}
\usepackage[unicode=true]{hyperref} 
\usepackage{listings}
\makeatletter

\newcommand{\pt}{\mathcal{PT}}

\DeclareMathOperator{\Tr}{Tr}

\hypersetup{colorlinks=true, linkcolor=blue, filecolor=blue, urlcolor=cyan,}
\makeatletter
\newcommand*\bigcdot{\mathpalette\bigcdot@{.5}}
\newcommand*\bigcdot@[2]{\mathbin{\vcenter{\hbox{\scalebox{#2}{$\m@th#1\bullet$}}}}}
\makeatother

\begin{document}

\title[Conserved quantities in non-Hermitian systems]
{Conserved quantities in non-Hermitian systems via vectorization method}


\author{Kaustubh S. Agarwal, Jacob Muldoon,Yogesh N. Joglekar}
\affiliation{Department of Physics, Indiana University-Purdue University Indianapolis, Indianapolis, Indiana 46202}

\begin{abstract}
Open classical and quantum systems have attracted great interest in the past two decades. These include systems described by non-Hermitian Hamiltonians with parity-time $(\mathcal{PT})$ symmetry that are best understood as systems with balanced, separated gain and loss. Here, we present an alternative way to characterize and derive conserved quantities, or intertwining operators, in such open systems.  As a consequence, we also obtain non-Hermitian or Hermitian operators whose expectations values show single exponential time dependence. By using a simple example of a $\mathcal{PT}$-symmetric dimer that arises in two distinct physical realizations, we demonstrate our procedure for static Hamiltonians and generalize it to time-periodic (Floquet) cases  where intertwining operators are stroboscopically conserved. Inspired by the Lindblad density matrix equation, our approach provides a useful addition to the well-established methods for characterizing time-invariants in non-Hermitian systems. 
\end{abstract}

\maketitle


\section{Introduction}
\label{sec:intro}

Since the seminal discovery of Bender and coworkers in 1998~\cite{Bender1998}, non-Hermitian Hamiltonians $H$ with real spectra have become a subject of intense scrutiny~\cite{levai2000,Bender2003,Bender2007}. The initial work on this subject focused on taking advantage of the reality of the spectrum to define a complex extension of quantum theory~\cite{Bender2002} where the traditional Dirac inner product is replaced by a Hamiltonian-dependent ($\mathcal{CPT}$) inner product. Soon it became clear that this process can be thought of as identifying positive definite operators $\hat{\eta}\geq 0$ that intertwine with the Hamiltonian~\cite{Mostafazadeh2002,Mostafazadeh2003,Mostafazadeh2010}, i.e. $\hat{\eta} H=H^\dagger\hat{\eta}$, and that a non-unique complex extension of standard quantum theory is generated by each positive definite $\eta$~\cite{Znojil2006,Znojil2009}.  These mathematical developments were instrumental to elucidating the role played by non-Hermitian, self-adjoint operators, biorthogonal bases, and non-unitary similarity transformations that change an orthonormal basis set into a non-orthogonal, but linearly independent basis set in physically realizable classical and quantum models~\cite{Znojil2015}. 

A decade later, this mathematical approach gave way to experiments with the recognition that non-Hermitian Hamiltonians that are invariant under combined operations of parity and time-reversal ($\mathcal{PT}$) represent open systems with balanced gain and loss~\cite{Ruschhaupt2005,ElGanainy2007a,ElGanainy2008,Klaiman2008}.  The spectrum of a $\mathcal{PT}$-symmetric Hamiltonian $H_\text{PT}(\gamma)$ is purely real when the non-Hermiticity $\gamma$ is small. With increasing $\gamma$, a level attraction and resulting degeneracy turns the spectrum into complex-conjugate pairs when the non-Hermiticity exceeds a nonzero threshold $\gamma_\text{PT}$~\cite{Joglekar2013}. This transition is called $\mathcal{PT}$-symmetry breaking transition, and at the threshold $\gamma_\text{PT}$ the algebraic multiplicity of the degenerate eigenvalue is larger than the geometric multiplicity, i.e. an exceptional point (EP)~\cite{Kato1995}. 

Fueled by this physical insight, the past decade has seen an explosion of experimental platforms, usually in classical wave systems, where effective $\mathcal{PT}$-symmetric Hamiltonians with balanced gain and loss have been realized. They include evanescently coupled waveguides~\cite{Ruter2010}, fiber loops~\cite{Regensburger2012}, microring resonators~\cite{Hodaei2014,Peng2014}, optical resonators ~\cite{Chang2014}, electrical circuits ~\cite{Schindler2011,Wang2020,Quirozjuarez2021}, and mechanical oscillators~\cite{Bender2013}. The key characteristics of this transition, driven by the non-orthogonality of eigenstates, are also seen in systems with mode-selective losses~\cite{Duchesne2009,LeonMontiel2018,Joglekar2018}. In the past two years, these ideas have been further extended to minimal quantum systems, thereby leading to observation of $\mathcal{PT}$-symmetric breaking and attendant phenomena in a single spin~\cite{Wu2019}, a single superconducting transmon~\cite{naghiloo2019}, ultracold atoms~\cite{Li2019}, and quantum photonics~\cite{klauck2019}. 

We remind the readers the effective Hamiltonian approach requires Dirac inner product, and is valid in both $\mathcal{PT}$-symmetric and $\mathcal{PT}$-broken regions. Apropos, the non-unitary time evolution generated by the effective $H_\text{PT}$ signals the fact that the system under consideration is open. In this context, every intertwining operator $\hat{\eta}$---positive definite or not---represents a time-invariant of the system. In other words, although the state norm $\langle\psi(t)|\psi(t)\rangle$ or the energy $\langle\psi(t)|H_\text{PT}|\psi(t)\rangle$ of a state $|\psi(t)\rangle=\exp(-iH_\text{PT}t)|\psi(0)\rangle$ of a $\mathcal{PT}$-symmetric system are not conserved~\cite{Mostafazadeh2010}, the expectation values $\langle\psi(t)|\hat{\eta}|\psi(t)\rangle$ remain constant with time. For a system with $N$ degrees of freedom, a complete characterization of intertwining operators for a given system is carried out by solving the set of $N^2$ simultaneous, linear equations, i.e. 
\begin{align}
\label{eq:int}
\hat{\eta}H_\text{PT}=H_\text{PT}^\dagger\hat{\eta}.
\end{align}

In the past, several different avenues have been used to obtain these conserved quantities. They include spectral decomposition methods~\cite{Mostafazadeh2010,Ruzicka2021}, an explicit recursive construction to generate a tower of intertwining operators~\cite{Bian2020,Quirozjuarez2021}, sum-rules method~\cite{Berry2008}, and the  Stokes parametrization approach for a $\mathcal{PT}$-symmetric dimer~\cite{Teimourpour2014}. Here, we present yet another approach to the problem, and illustrate it with two simple examples. The plan of the paper is as follows. In Sec.~\ref{sec:lindblad}, we present the eigenvalue-equation approach for  intertwining operators and the details of the vectorization scheme. This method is valid for any finite dimensional $\mathcal{PT}$-symmetric Hamiltonian. In Sec.~\ref{sec:qd}, we present results of such analysis for a quantum $\mathcal{PT}$-symmetric dimer with static or time-periodic gain and loss. Corresponding results for a classical $\mathcal{PT}$-symmetric dimer are presented in Sec.~\ref{sec:cd}. We conclude the paper with a brief discussion in Sec.~\ref{sec:conc}. 


\section{Intertwining operators as an eigenvalue problem}
\label{sec:lindblad}
For a $\mathcal{PT}$-symmetric system undergoing coherent but non-unitary dynamics with static Hamiltonian $H_\text{PT}$, the expectation value of an operator $\hat{\eta}$ satisfies the following linear-in-$\hat{\eta}$ first-order differential equation
\begin{align}
\label{eq:evproblem}
\partial_t\langle\psi(t)|\hat{\eta}|\psi(t)\rangle=-i\langle\psi(t)|\hat{\eta}H_\text{PT}-H^\dagger_\text{PT}\hat{\eta}|\psi(t)\rangle. 
\end{align}
This equation is reminiscent of the Gorini Kossakowski Sudarshan Lindblad (GKSL) equation
~\cite{Gorini1976,Lindblad1976} (henceforth referred to as the Lindblad equation) that describes the dynamics of the reduced density matrix of a quantum system coupled to a much larger environment
~\cite{Ban1993,Albert2014,Manzano2020}. Interpreting $\hat{\eta}$ as an $N\times N$ matrix, all $\hat{\eta}$s that satisfy Eq.(\ref{eq:evproblem}) can be obtained from the corresponding eigenvalue problem 
\begin{align}
\label{eq:ev2}
\mathcal{E}_k\hat{\eta}_k=-i(\hat{\eta}_k H_\text{PT}- H^\dagger_\text{PT}\hat{\eta}_k)\equiv\mathcal{L}\hat{\eta}_k,
\end{align}
 for $1\leq k\leq N^2$. We vectorize the matrix $\hat{\eta}$ into an $N^2$-sized column vector $|\eta^v\rangle$ by stacking its columns, i.e. $[\hat{\eta}]_{pq}\rightarrow \eta^v_{p+(q-1)N}$~\cite{Gunderson2021}. Under this vectorization, the Hilbert-Schmidt trace inner product carries over to the Dirac inner product, $\Tr(\hat{\eta}_1^\dagger\hat{\eta}_2)=\langle \eta^v_1|\eta^v_2\rangle$ where $\langle\eta^v_1|$ is the Hermitian-conjugate row vector obtained from the column vector $|\eta^v_1\rangle$. Using the identity $A\hat{\eta}B\rightarrow (B^T\otimes A)|\eta^v\rangle$, the eigenvalue problem Eq. (\ref{eq:ev2}) becomes $\det(\mathcal{L}-\mathcal{E}\mathbbm{1}_{N^2})=0$ where the $N^2\times N^2$ ``Liouvillian'' matrix is given by 
\begin{align}
\label{eq:L}
\mathcal{L}=-i\left[H_\text{PT}^T\otimes\mathbbm{1}_N-\mathbbm{1}_N\otimes H_\text{PT}^\dagger\right], 
\end{align}
and $\mathbbm{1}_m$ is the $m\times  m$ identity matrix. Thus, the intertwining operators are distinct eigenvectors $|\eta^v_m\rangle$ with zero eigenvalue in Eq.(\ref{eq:ev2}). The $N^2$ eigenvalues of the Liouvillian $\mathcal{L}$ are simply related to $N$ eigenvalues $\epsilon_m$ of the $H_\text{PT}$ as 
\begin{align}
\label{eq:lambda}
\mathcal{E}_{pq}=-i(\epsilon_p-\epsilon_q^*).
\end{align}
Since the spectrum of $H_\text{PT}$ is either real ($\epsilon_p=\epsilon_p^*$) or complex conjugates ($\epsilon_p=\epsilon_q^*$ for some pair), there are $N$ zero eigenvalues of $\mathcal{L}$ when $H_\text{PT}$ has no symmetry-driven degeneracies; the number of zero eigenvalues grows to $N^2$ if the Hamiltonian is proportional to the identity matrix~\cite{Ruzicka2021}. This analysis also provides a transparent way to construct corresponding intertwining operators via the spectral decomposition of $H_\text{PT}$~\cite{Mostafazadeh2010}. Note that when $\mathcal{E}=0$, due to the linearity of the intertwining relation, Eq.(\ref{eq:int}), without loss of generality, we can choose the $N$ intertwining operators $\hat{\eta}_m$ to be Hermitian. 

So what is the advantage of this approach? For one, it gives us $N(N-1)$ other, (generally non-Hermitian)  operators whose expectation value {\it in any arbitrary state evolves simply exponentially in time.} When $\mathcal{E}_{pq}$ is purely imaginary, it leads to the non-Hermitian $\hat{\eta}_{pq}$ whose expectation value in any state remains constant in magnitude; on the other hand, if $\mathcal{E}_{pq}$ is purely real, one can choose a Hermitian $\hat{\eta}_{pq}$ whose expectation value exponentially grows or decays with time. 

This analysis of constants of motion is valid for systems with a static, $\mathcal{PT}$-symmetric Hamiltonian. It can be suitably generalized to time-periodic, $\mathcal{PT}$-symmetric Hamiltonians via the Floquet formalism~\cite{Joglekar2014,Lee2015,Li2019,LeonMontiel2018,Quirozjuarez2021}. When $H_\text{PT}(t)=H_\text{PT}(t+T)$ is periodic in time, the long-time dynamics of the system is governed by the Floquet time-evolution operator~\cite{Hanggi1998}
\begin{align}
\label{eq:gf}
G_F(T) = \mathbbm{T}e^{-i\int_0^T H_\text{PT}(t') dt'},
\end{align}
where $\mathbbm{T}$ stands for the time ordered product that takes into account non-commuting nature of the Hamiltonians at different times. The (stroboscopic) dynamics of the system at times $t_m=mT$ is then given by $|\psi(t_m)\rangle=G_F^m|\psi(0)\rangle$, and the corresponding, 
Hermitian, conserved operators $\hat{\eta}=\hat{\eta}^\dagger$ are determined by~\cite{Ruzicka2021,Quirozjuarez2021}
\begin{align}
\label{eq:int2}
G_F^\dagger\hat{\eta}G_F=\hat{\eta}.
\end{align}
Vectorization of Eq.(\ref{eq:int2}) implies the conserved quantities are given by eigenvectors of the ``Floquet Liouville time-evolution''  matrix 
\begin{align}
\label{eq:G}
\mathcal{G}=G_F^T\otimes G_F^\dagger
\end{align}
 with unit eigenvalue. Since $G_F(T)$ inherits the $\mathcal{PT}$ symmetry of the time-periodic Hamiltonian, the eigenvalues $\kappa_m$ of $G_F(T)$ either lie on a circle ($|\kappa_p|=\text{const.}$; $\mathcal{PT}$-symmetric phase) or occur along a radial line in pairs with constant geometric mean ($|\kappa_p\kappa_q|=\text{const.}$; $\mathcal{PT}$-broken phase). Therefore, it is straightforward to see that among the $N^2$ eigenvalues $\lambda_{pq}\equiv \kappa_p\kappa_q^*$ of $\mathcal{G}$, there are $N$ unit eigenvalues, giving rise to $N$ conserved quantities. As in the case with the static Hamiltonian, the remaining $N(N-1)$ eigenvectors give operators that vary exponentially with the stroboscopic time $t_m$ irrespective of the initial state $|\psi(0)\rangle$. If $\lambda_{pq}$ is real, we can choose them to be Hermitian, as in the case of a static Hamiltonian.

We now demonstrate these ideas with two concrete examples. 


\section{Quantum $\mathcal{PT}$-symmetric dimer}
\label{sec:qd}

We first consider the prototypical $\mathcal{PT}$-symmetric dimer ($N=2$) with a Hamiltonian given by 
\begin{align}
\label{eq:h1}
H_1(t) =J\sigma_x + i\gamma f(t)\sigma_z =H_1^T\neq H^\dagger_1. 
\end{align}
We call this model ``quantum'' because it arises naturally in minimal quantum systems undergoing Lindblad evolution when we confine ourselves to trajectories that undergo no quantum jumps~\cite{naghiloo2019}, as well as in wave systems~\cite{Ruter2010,Regensburger2012,Hodaei2014,Peng2014,Chang2014}. Here $J>0$ denotes coupling between the two degrees of freedom and $\gamma>0$ is the strength of the gain-loss term. $H_1$ is $\mathcal{PT}$-symmetric with the parity operator $\mathcal{P}=\sigma_x$ and time-reversal operator $\mathcal{T}=*$ (complex conjugation). The eigenvalues $\epsilon_{1,2}=\pm\sqrt{J^2-\gamma^2}\equiv\pm\Delta(\gamma)$ of the Hamiltonian $H_1(\gamma)$ remain real when $\gamma<\gamma_\text{PT}=J$ and become purely imaginary when $\gamma$ exceeds the threshold. 

In the static case, $f(t)=1$, using Eq.~\ref{eq:int}, it is easy to show that $\hat{\eta}_1=\mathcal{P}=\sigma_x$ is the first intertwining operator~\cite{Bian2020,Ruzicka2021}, and the recursive construction gives the second intertwining operator as $\hat{\eta}_2=\hat{\eta}_1 H_1/J=\mathbbm{1}+(\gamma/J)\sigma_y$. However, the corresponding $4\times 4$ Liouvillian matrix $\mathcal{L}$, Eq.(\ref{eq:L}), has two nonzero eigenvalues that are given by  $\mathcal{E}_\pm=\pm 2i\Delta$. The corresponding eigen-operators are given by
\begin{align}
\label{eq:etapm}
\hat{\eta}_\pm=\frac{1}{J^2}\left[\begin{array}{cc}
(\gamma\pm i\Delta)^2 & -i(\gamma\pm i\Delta)\\
+i(\gamma\pm i\Delta) & 1
\end{array}\right].
\end{align}
Note that the $2\times 2$ matrices $\hat{\eta}_\pm$ have rank 1, and thus are not invertible. In the $\mathcal{PT}$-symmetric region ($\Delta\in\mathbb{R}$), the operators $\hat{\eta}_\pm$ are not Hermitian, whereas in the $\mathcal{PT}$ broken region ($\Delta\in i\mathbbm{R}$), they are Hermitian. 

Next we consider the time-periodic case, i.e. $f(t)=f(t+T)$ where $f(t)=\text{sgn}(t)$ for $|t|<T/2$ denotes a square wave. This piecewise constant gain and loss means that the Hamiltonian switches from $H_{1+}=J\sigma_x+i\gamma\sigma_z$ for $0\leq t< T/2$ to $H_{1-}=\mathcal{T}H_{1+}\mathcal{T}=J\sigma_x-i\gamma\sigma_z$ for $T/2\leq t<T$.  The non-unitary Floquet time-evolution operator can be explicitly evaluated as~\cite{Harter2020}
\begin{align}
\label{eq:GF1}
G_F(T)&= e^{-iH_{1-}T/2}e^{-iH_{1+}T/2},\\
\label{eq:GF2}
&=G_0\mathbbm{1}_2+iG_x\sigma_x+G_y\sigma_y,
\end{align}
where $G_0=[J^2\cos(\Delta T)-\gamma^2]/\Delta^2$, $G_x=-J\sin(\Delta T)/\Delta$ and $G_y=-J\gamma[1-\cos(\Delta T)]/\Delta^2$ are coefficients that remain real irrespective of where $\Delta(\gamma)$ is real or purely imaginary. When $\gamma\rightarrow 0$, this reproduces the expected result $G_F(T)=\exp(-iJ\sigma_x T)$ and in the limit $T\rightarrow 0$, the time-evolution operator reduces to $\mathbbm{1}_2$ as expected. On the other hand, as $\Delta\rightarrow 0$, the power series for $G_F(T) $ terminates at {\it second order in $T$} in a sharp contrast to the static case, where it terminates at first order in time. 

The eigenvalues of $G_F$, Eq.(\ref{eq:GF2}), are 
\begin{align}
\kappa_{1,2}=G_0\pm i\sqrt{G_x^2-G_y^2}. 
\end{align}
Thus the EP contours separating the $\mathcal{PT}$-symmetric phase ($|\kappa_1|=|\kappa_2|$) from the $\mathcal{PT}$-broken phase ($|\kappa_1|\neq|\kappa_2|$) are given by $G_x=\pm G_y$~\cite{Harter2020}.  It is easy to check that $\hat{\eta}_1=\sigma_x$ satisfies $G_F^\dagger\hat{\eta}_1G_F=\hat{\eta}_1$ and is a stroboscopically conserved quantity. The second conserved operator is obtained from the symmetrized or antisymmetrized version of the recursive construction~\cite{Ruzicka2021}, i.e. 
\begin{align}
\label{eq:eta2}
\hat{\eta}_2=\left\{\begin{array}{c} 
(\hat{\eta}_1 G_F+ G_F^\dagger\hat{\eta}_1)/2,\\ 
-i(\hat{\eta}_1 G_F- G_F^\dagger\hat{\eta}_1)/2.
\end{array}\right.
\end{align}
In the present case, the symmetrized version returns $\hat{\eta}_1$ while the antisymmetrized version gives the second, linearly independent conserved operator as $\hat{\eta}_2=G_x\mathbbm{1}_2+G_y\sigma_z$. Following the procedure outlined in Sec.~\ref{sec:lindblad} gives us two unity eigenvalues of $\mathcal{G}$, Eq.(\ref{eq:G}), with corresponding conserved operators. The remaining two eigenvalues are complex conjugates with unit length in the $\mathcal{PT}$-symmetric region, i.e. $\lambda_3=\lambda_4^*=e^{i\phi}$ with eigen-operators $\hat{\eta}_+=\hat{\eta}_{-}^\dagger$ that are Hermitian conjugates of each other. In the $\mathcal{PT}$-broken region, the two complex eigenvalues with equal phase satisfy $|\lambda_3\lambda_4|=1$. 

\begin{figure*}
\centering
\includegraphics[width=\textwidth]{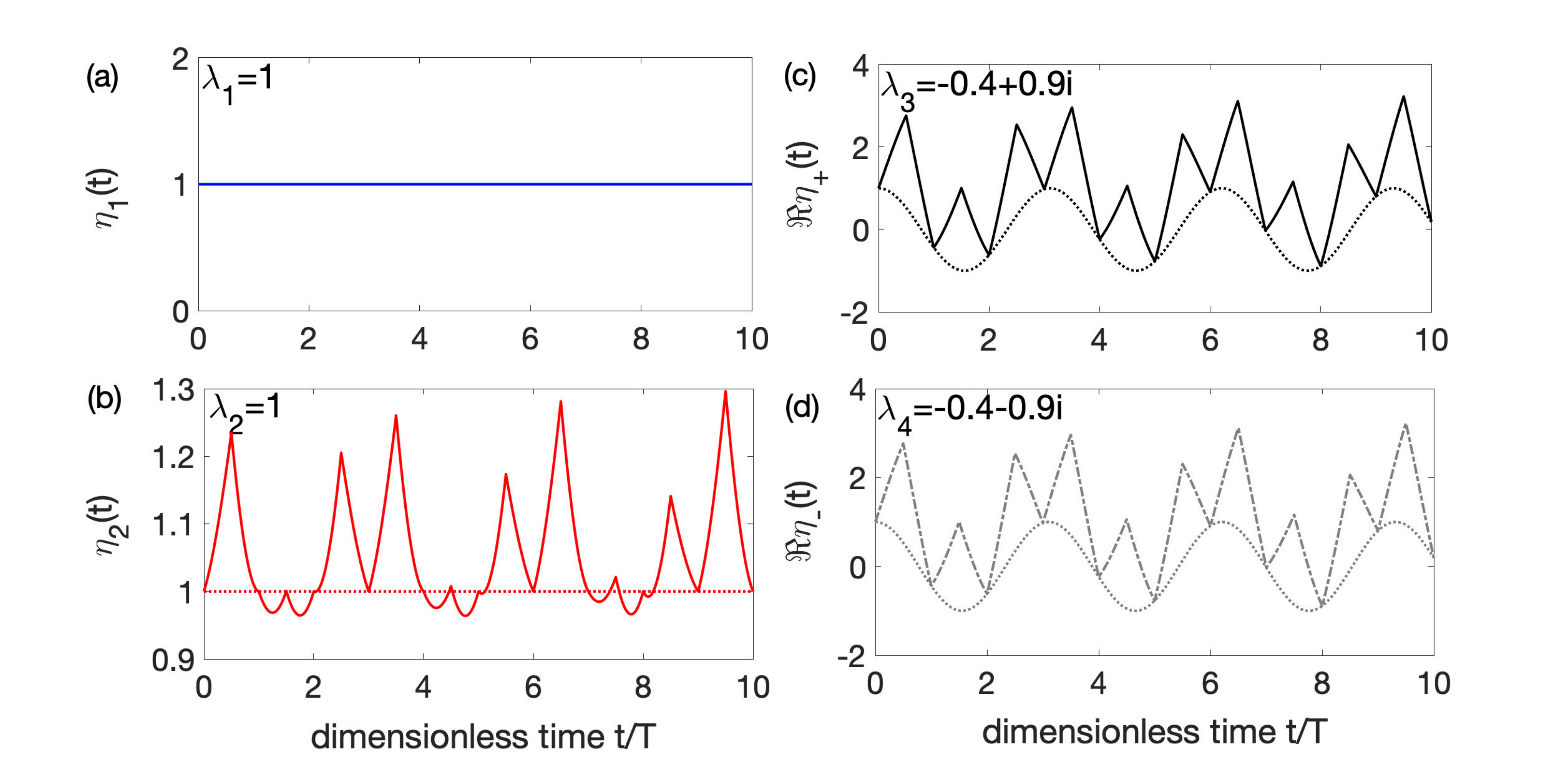}
\caption{Conserved quantities for a Floquet, quantum $\mathcal{PT}$-symmetric dimer. System parameters are $\gamma=0.5J$, $JT=1$, $|\psi(0)\rangle=|+x\rangle$, and $\eta_\alpha(t)$ denote normalized expectation values. (a) $\hat{\eta}_1=\sigma_x$ is an eigen-operator of $\mathcal{G}$ with eigenvalue $\lambda_1=1$; $\eta_1(t)$ is constant. (b) $\hat{\eta}_2=G_x\mathbbm{1}_2+G_y\sigma_z$ is the second eigen-operator of $\mathcal{G}$ with $\lambda_2=1$; $\eta_2(t)$ oscillates with time, but is stroboscopically constant at $t/T=n$; the dotted red line shows $\Re\lambda_2^t=1$. (c) $\hat{\eta}_+$ is a non-Hermitian eigen-operator with unit-length eigenvalue $\lambda_3=-0.44+0.9i$. The real part of its normalized expectation value stroboscopically matches $\Re\lambda_3^t$ shown in dotted black. (d) Corresponding result for $\hat{\eta}_{-}=\hat{\eta}_+^\dagger$ with eigenvalue $\lambda_4=\lambda_3^*$.} 
\label{fig:qd}
\end{figure*}

Figure~\ref{fig:qd} shows expectation values normalized to their initial values, 
\begin{align}
\label{eq:normeta}
\eta_\alpha(t)\equiv \frac{\langle\psi(t)|\hat{\eta}_\alpha|\psi(t)\rangle}{\langle\psi(0)|\hat{\eta}_\alpha|\psi(0)\rangle}
\end{align}
calculated with initial state $|\psi(0)\rangle=|+x\rangle$ as a function of dimensionless time $t/T$. The system parameters are $\gamma=0.5J$, $JT=1$, and $|+x\rangle$ is the eigenstate of $\sigma_x$ with eigenvalue +1. Thus, the system is in the $\mathcal{PT}$-symmetric region.  Figure~\ref{fig:qd}a shows that $\eta_1(t)$ is conserved in this evolution at all times, not just stroboscopically at $t_m=mT$. On the other hand $\eta_2(t)$, shown in Fig.~\ref{fig:qd}b, has a periodic behavior with a period $\sim 30T$ (not shown). Although $\eta_2(t)$ varies with time, it is stroboscopically conserved, $\eta_2(t_m)=1$. The dotted red line shows $\Re\lambda_2^t=1$. Figure~\ref{fig:qd}c shows that the real part of $\eta_{+}(t)$, with eigenvalue $\lambda_3=-0.44+0.9i$, also shows periodic variation. The dotted black line shows $\Re\lambda_3^t$, and the fact that $\Re\eta_{+}(t_m)$ matches it stroboscopically confirms the simple sinousoidal variation of this eigen-operator. Figure~\ref{fig:qd}d shows corresponding results for the fourth operator $\hat{\eta}_{-}=\hat{\eta}_{+}^\dagger$ with eigenvalue $\lambda_4=-0.44+0.9i$. 

We conclude this section with transformation properties of $G_F(T)$ and the conserved operators $\hat{\eta}$. When the periodic Hamiltonian is Hermitian, i.e. $H_0(t)=H_0^\dagger(t)=H_0(t+T)$, shifting the zero of time to $t_0$ leads to a unitary transformation,
\begin{align}
\label{eq:GF3}
G_F(T+t_0,t_0)& = U(t_0) G_F(T)U^\dagger(t_0),\\
U(t_0)&=\mathbbm{T}e^{-i\int_0^{t_0} H_0(t')dt'}. 
\end{align}
Therefore the conserved operators are also unitarily transformed. However, in our case, Eq.(\ref{eq:GF3}) becomes a {\it similarity transformation}, $G_F(T+t_0,t_0)=SG_F(T)S^{-1}$ where $S=\mathbbm{T}\exp(-i\int_{0}^{t_0}H_\text{PT}(t')dt')$ does not satisfy $S^\dagger S=\mathbbm{1}=SS^\dagger$. Under this transformation, the conserved operators change as $\hat{\eta}\rightarrow S^{-1\dagger}\hat{\eta}S^{-1}$. This non-unitary transformation of the conserved quantities under a shift of zero of time suggests that they are not related to ``symmetries'' of the open system with balanced gain and loss.  


\section{Classical $\mathcal{PT}$-symmetric dimer}
\label{sec:cd}

We now consider a different example characterized by a non-Hermitian Hamiltonian with purely imaginary entries. We call such a system ``classical'' because having $H_\text{PT}=-H^*_\text{PT}$ ensures that the non-unitary time evolution operator $\exp(-iH_\text{PT}t)$ is purely real, and therefore $|\psi(t)\rangle$ remains real if $|\psi(0)\rangle$ is. Such classical Hamiltonian arises naturally in describing the energy density dynamics in mechanical or electrical circuits~\cite{Schindler2011,Bender2013,LeonMontiel2018,Wang2020,Quirozjuarez2021}, where $|\psi(t)\rangle$ encodes time-dependent positions, velocities, voltages, currents, etc. and is obviously real. As its simplest model, we consider a dimer governed by the Hamiltonian 
\begin{equation}
\label{eq:h2}
H_2(t) = J\sigma_y + i\gamma f(t) \sigma_z=-H_2^*.
\end{equation}

On one level, the Hamiltonian $H_2(t)$, Eq.(\ref{eq:h2}), is ``just a change of basis'' from $H_1(t)$, Eq.(\ref{eq:h1}); $H_2(t)=\exp(-i\pi\sigma_z/4)H_1(t)\exp(+i\pi\sigma_z/4)$. However, since $H_2(t)$ models effective, classical systems where the entire complex state space is physically accessible, it is necessary to treat it differently. A physical realization of $H_2(t)$ is found in a single LC circuit whose inductance $L(t)$ and capacitance $C(t)$ are varied such that its characteristic frequency $J=1/\sqrt{L(t)C(t)}$ remains constant~\cite{Quirozjuarez2021}. 

Hamiltonian $H_2(t)$ is $\pt$-symmetric with $\mathcal{PT}=\sigma_x *$. In the static case ($f(t)=1$), the two, Hermitian intertwining operators are given by $\hat{\eta}_1=\sigma_y$ and $\hat{\eta}_2=\hat{\eta}_1H_2/J=\mathbbm{1}_2-(\gamma/J)\sigma_x$. In addition, the vectorization approach gives two, rank-1 eigen-operators
\begin{align}
\label{eq:etapm2}
\hat{\eta}_\pm=\frac{1}{J^2}\left[\begin{array}{cc}
(\gamma\pm i\Delta)^2 & -(\gamma\pm i\Delta)\\
-(\gamma\pm i\Delta) & 1
\end{array}\right],  
\end{align}
with eigenvalues $\mathcal{E}_\pm=\pm2i\Delta$. As we discussed in Sec.~\ref{sec:qd}, these operators are not Hermitian in the $\mathcal{PT}$-symmetric phase, and become Hermitian in the $\mathcal{PT}$ broken phase. 

For the Floquet case, we choose a gain-loss term that is nonzero only at discrete times. This is accomplished by choosing the dimensionless function $f(t)$ as 
\begin{align}
\label{eq:ft}
f(t)=T\left[\delta(t)-\delta(t-T/2)\right]=f(t+T).
\end{align}
The resulting Floquet time-evolution operator $G_F(T)$ can be analytically calculated~\cite{Quirozjuarez2021}. Since the Hamiltonian $H_2(t)$ is Hermitian at all times except $t_k=kT/2$, the evolution is mostly unitary, punctuated by non-unitary contributions that occur due to $\delta$-functions at times $t_k$. The result is 
\begin{align}
\label{eq:Gf2}
G_F(T)&= e^{+\gamma T\sigma_z}e^{-iJT\sigma_y/2}e^{-\gamma T\sigma_z}e^{-iJT\sigma_y/2}\nonumber\\
&=G_0\mathbbm{1}_2+G_x\sigma_x+iG_y\sigma_y+G_z\sigma_z,
\end{align}
where the four real coefficients $G_k$ are given by 
\begin{align}
\label{eq:g0}
G_0&=\cos^2(JT/2)-\sin^2(JT/2)\cosh(2\gamma T),\\
\label{eq:gx}
G_x&=-\sin(JT)\sinh(2\gamma T)/2,\\
\label{eq:gy}
G_y&=-\sin(JT)[1+\cosh(2\gamma T)]/2.\\
\label{eq:gz}
G_z&=-\sin^2(JT/2)\sinh(2\gamma T).
\end{align}
As is expected, the purely real $G_F(T)$ reduces to $\exp(-iJT\sigma_y)$ in the Hermitian limit $\gamma\rightarrow 0$. The EP conoturs, on the other hand, are determined by the constraint $G_x^2+G_y^2-G_z^2=0$, which reduces to $\cos(JT/2)=\tanh(\gamma T)$~\cite{Quirozjuarez2021}.

Two linearly independent Floquet intertwining operators obtained by solving Eq.(\ref{eq:int2}) are given by $\hat{\eta}_1=\sigma_y$ and $\hat{\eta}_2=-i(\hat{\eta}_1G_F-G_F^\dagger\hat{\eta}_1)/2$. The latter simplifies to $\hat{\eta}_2=G_y\mathbbm{1}_2+G_z\sigma_x-G_x\sigma_z$. We leave it for the reader to check that, as in the case of Floquet quantum $\mathcal{PT}$ dimer problem, the symmetrized version of the recursive procedure, Eq.(\ref{eq:eta2}), does not lead to a result that is linearly independent of $\hat{\eta}_1$. Following the recipe in Sec.~\ref{sec:lindblad}, we supplement these analytical results with symbolic or numerical results for four eigenvalues $\lambda_k$ and four eigen-operators $\hat{\eta}_1,\hat{\eta}_2,\hat{\eta}_\pm$ of $\mathcal{G}$, Eq.(\ref{eq:G}).

\begin{figure}
\centering
\includegraphics[width=\columnwidth]{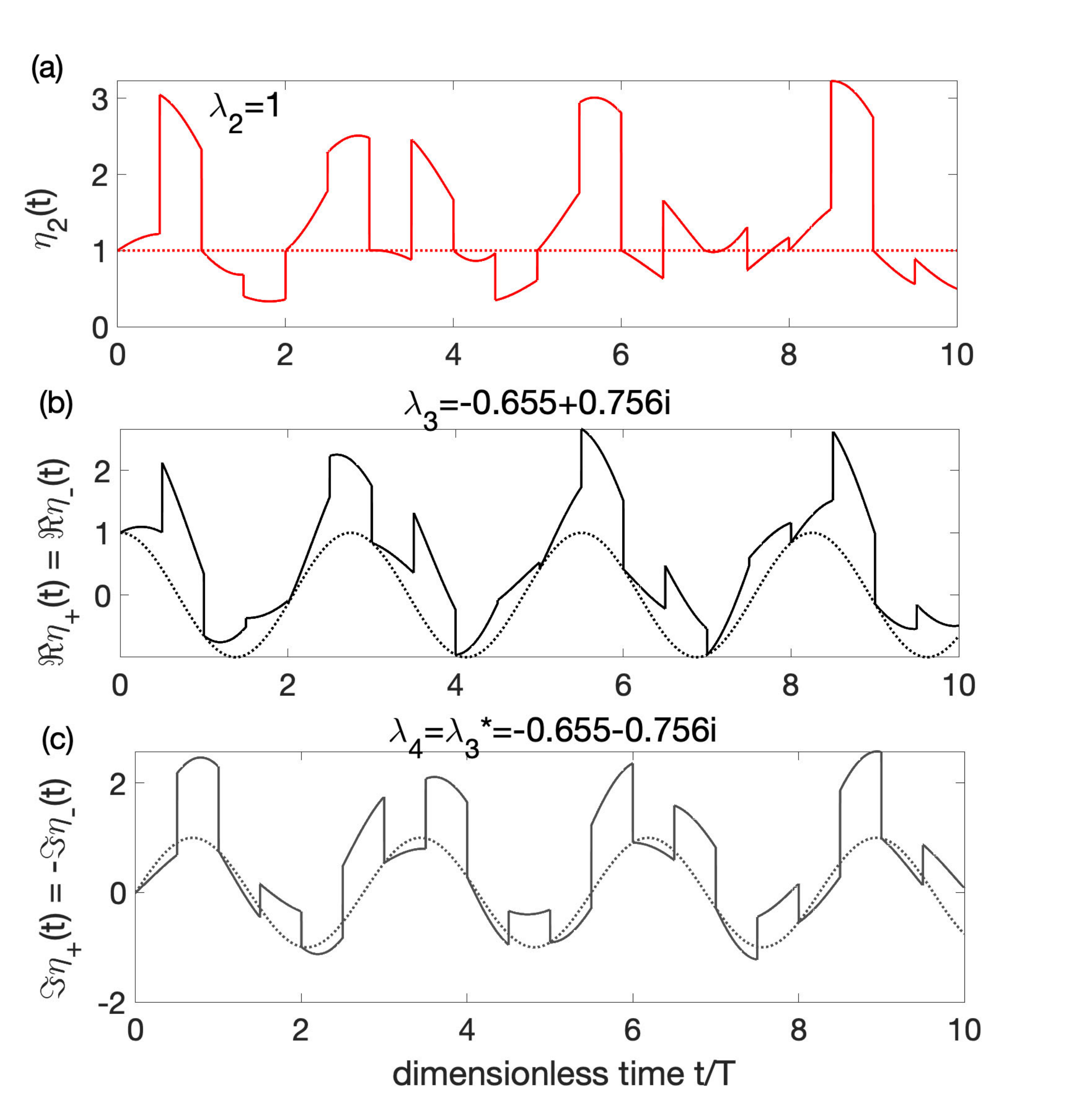}
\caption{Conserved quantities for a classical  $\mathcal{PT}$-symmetric dimer with $\gamma=0.5J$, $JT=1$, $|\psi(0)\rangle=|+x\rangle$. Since $|\psi(t)\rangle$ is purely real, the expectation value of $\hat{\eta}_1=\sigma_y$ is always zero. (a) $\hat{\eta}_2=G_y\mathbbm{1}_2+G_x\sigma_z-G_z\sigma_x$ is the second eigen-operator of $\mathcal{G}$ with $\lambda_2=1$. $\eta_2(t)$ oscillates with time, but is stroboscopically constant at $t/T=n$; the dotted red line shows $\Re\lambda_2^t=1$. (b) Since the system is in the $\mathcal{PT}$-symmetric phase, $\Re\eta_{+}(t)=\Re\eta_{-}(t)$ (solid black) shows periodic behavior with values that stroboscopically match $\Re\lambda_3^t$, shown in dotted black. (c) Corresponding imaginary parts, $\Im\eta_{-}(t)=-\Im\eta_+(t)$ (dot-dashed black) show similar, stroboscopically matching behavior.} 
\label{fig:cd}
\end{figure}

Figure~\ref{fig:cd} shows the behavior of normalized expectation values $\eta_\alpha(t)$ calculated with $|\psi(0)\rangle=|+x\rangle$ as a function of time. The system parameters are $\gamma=0.5J$ and  $JT=1$, and therefore the system is in the $\mathcal{PT}$-symmetric region.  Note that since $|\psi(t)\rangle$ is purely real,  $\langle\psi(t)|\hat{\eta}_1|\psi(t)\rangle=0$  independent of time~\cite{Quirozjuarez2021}. On the other hand $\eta_2(t)$, shown in Fig.~\ref{fig:cd}a, has a periodic behavior. Although $\eta_2(t)$ varies with time, it is stroboscopically conserved, $\eta_2(t_m)=1$.  Figure~\ref{fig:cd}b shows that the real part of $\eta_{+}(t)$, with unit-magnitude eigenvalue $\lambda_3=-0.65+0.756i$, also varies periodically. The dotted black line shows $\Re\lambda_3^t$, and the fact that $\Re\eta_{+}(t_m)$ matches it stroboscopically confirms the simple sinousoidal variation of this eigen-operator. Since the system is in the $\mathcal{PT}$-symmetric phase, $\hat{\eta}_{-}=\hat{\eta}_+^\dagger$, and therefore $\Re\eta_{-}(t)=\Re\eta_+(t)$. Figure~\ref{fig:cd}c shows the corresponding imaginary parts $\Im\eta_{+}(t)=-\Im\eta_{-}(t)$ for the eigen-operator with the complex conjugate eigenvalue $\lambda_4=\lambda_3^*$. We note that in the $\mathcal{PT}$-broken regime, the non-unit-modulus eigenvalues are not complex conjugates of each other, and therefore the corresponding eigen-operators will not satisfy the relations shown in Figs.~\ref{fig:cd}b-c. 


\section{Conclusions}
\label{sec:conc}
In this article, we have presented a new method to obtain intertwining operators or conserved quantities in $\mathcal{PT}$-symmetric systems with static or time-periodic Hamiltonians. In this approach, these operators appear as zero-$\mathcal{E}$ eigenmodes of the static Liouvillian $\mathcal{L}$ or as $\lambda=1$  eigenmodes of the Floquet $\mathcal{G}$. For an $N$-dimensional system, in addition to the $N$ constants of motion, this approach also leads to $N(N-1)$ operators whose expectation values {\it in any arbitrary state} undergo simple exponential-in-time change. We have demonstrated these concepts with two simple, physically motivated examples of a $\mathcal{PT}$-symmetric dimer with different, periodic gain-loss profiles. We have deliberating stayed away from continuum models because 
extending this approach or the recursive construction~\cite{Bian2020,Ruzicka2021} to infinite dimensions will probably be plagued by challenges regarding domains of resulting, increasingly higher-order differential operators.  

The definition of an intertwining operator via Eq.(\ref{eq:int}) can be generalized to obtain conserved observables for Hamiltonians that posses other antilinear symmetries, such as anti-$\mathcal{PT}$ symmetry~\cite{Peng2016,Choi2018,Zhang2020} or anyonic-$\mathcal{PT}$ symmetry~\cite{Longhi2019,Arwas2021}. The recursive procedure to generate a tower of such operators~\cite{Ruzicka2021}, and the vectorization method presented in Sec.~\ref{sec:lindblad} remains valid for arbitrary antilinear symmetry. Thus, this approach can be used to investigate constants of motion in such systems as well.


\bibliographystyle{actapoly}
\bibliography{biblio}

\begin{thebibliography}{10}
\providecommand{\url}[1]{\texttt{#1}}
\providecommand{\urlprefix}{URL }
\providecommand{\eprint}[2][]{\url{#2}}
\makeatletter
\def\bibdoi{\begingroup\def\do##1{\catcode
  `##112\relax}\do$\do\\\do\_\do\%\do\^\expandafter\endgroup\@bibdoi}
\def\@bibdoi#1{\href{https://doi.org/#1}{\ttfamily{https://doi.org/#1}}}
\makeatother

\bibitem{Bender1998}
C.~M. Bender, S.~Boettcher.
\newblock Real spectra in non-hermitian hamiltonians having $\mathcal{PT}$
  symmetry.
\newblock \textit{Phys Rev Lett} \textbf{80}:5243--5246, 1998.
\newblock \bibdoi{10.1103/PhysRevLett.80.5243}.

\bibitem{levai2000}
G.~L\'evai, M.~Znojil.
\newblock Systematic search for pt-symmetric potentials with real energy
  spectra.
\newblock \textit{J Phys A: Math Gen} \textbf{33}(40):7165, 2000.

\bibitem{Bender2003}
C.~M. Bender, D.~C. Brody, H.~F. Jones.
\newblock Must a hamiltonian be hermitian?
\newblock \textit{American Journal of Physics} \textbf{71}(11):1095--1102,
  2003.
\newblock \bibdoi{10.1119/1.1574043}.

\bibitem{Bender2007}
C.~M. Bender.
\newblock Making sense of non-hermitian hamiltonians.
\newblock \textit{Reports on Progress in Physics} \textbf{70}(6):947--1018,
  2007.
\newblock \bibdoi{10.1088/0034-4885/70/6/r03}.

\bibitem{Bender2002}
C.~M. Bender, D.~C. Brody, H.~F. Jones.
\newblock {Complex Extension of Quantum Mechanics}.
\newblock \textit{Physical Review Letters} \textbf{89}(27):1--4, 2002.
\newblock \bibdoi{10.1103/PhysRevLett.89.270401}.

\bibitem{Mostafazadeh2002}
A.~Mostafazadeh.
\newblock Pseudo-hermiticity versus pt symmetry: The necessary condition for
  the reality of the spectrum of a non-hermitian hamiltonian.
\newblock \textit{Journal of Mathematical Physics} \textbf{43}(1):205--214,
  2002.
\newblock \bibdoi{10.1063/1.1418246}.

\bibitem{Mostafazadeh2003}
A.~Mostafazadeh.
\newblock {Exact PT-symmetry is equivalent to Hermiticity}.
\newblock \textit{Journal of Physics A: Mathematical and General}
  \textbf{36}(25):7081--7091, 2003.
\newblock \bibdoi{10.1088/0305-4470/36/25/312}.

\bibitem{Mostafazadeh2010}
A.~Mostafazadeh.
\newblock Pseudo-hermitian representation of quantum mechanics.
\newblock \textit{International Journal of Geometric Methods in Modern Physics}
  \textbf{07}(07):1191--1306, 2010.
\newblock \bibdoi{10.1142/s0219887810004816}.

\bibitem{Znojil2006}
M.~Znojil, H.~B. Geyer.
\newblock Construction of a unique metric in quasi-hermitian quantum mechanics:
  Nonexistence of the charge operator in a $2\times 2$ matrix model.
\newblock \textit{Physics Letters B} \textbf{640}(1-2):52--56, 2006.
\newblock \bibdoi{10.1016/j.physletb.2006.07.028}.

\bibitem{Znojil2009}
M.~Znojil.
\newblock Complete set of inner products for a discrete
  $\mathcal{PT}$-symmetric square-well hamiltonian.
\newblock \textit{Journal of Mathematical Physics} \textbf{50}(12):122105,
  2009.
\newblock \bibdoi{10.1063/1.3272002}.

\bibitem{Znojil2015}
M.~Znojil.
\newblock Special issue {\textquotedblleft}pseudo-hermitian hamiltonians in
  quantum physics in 2014{\textquotedblright}.
\newblock \textit{International Journal of Theoretical Physics}
  \textbf{54}(11):3867--3870, 2015.
\newblock \bibdoi{10.1007/s10773-014-2501-2}.

\bibitem{Ruschhaupt2005}
A.~Ruschhaupt, F.~Delgado, J.~G. Muga.
\newblock Physical realization of -symmetric potential scattering in a planar
  slab waveguide.
\newblock \textit{Journal of Physics A: Mathematical and General}
  \textbf{38}(9):L171--L176, 2005.
\newblock \bibdoi{10.1088/0305-4470/38/9/l03}.

\bibitem{ElGanainy2007a}
R.~El-Ganainy, K.~G. Makris, D.~N. Christodoulides, Z.~H. Musslimani.
\newblock Theory of coupled optical pt-symmetric structures.
\newblock \textit{Opt Lett} \textbf{32}(17):2632--2634, 2007.
\newblock \bibdoi{10.1364/OL.32.002632}.

\bibitem{ElGanainy2008}
K.~G. Makris, R.~El-Ganainy, D.~N. Christodoulides, Z.~H. Musslimani.
\newblock Beam dynamics in $\mathcal{P}\mathcal{T}$ symmetric optical lattices.
\newblock \textit{Phys Rev Lett} \textbf{100}:103904, 2008.
\newblock \bibdoi{10.1103/PhysRevLett.100.103904}.

\bibitem{Klaiman2008}
S.~Klaiman, U.~G\"unther, N.~Moiseyev.
\newblock Visualization of branch points in $\mathcal{P}\mathcal{T}$-symmetric
  waveguides.
\newblock \textit{Phys Rev Lett} \textbf{101}:080402, 2008.
\newblock \bibdoi{10.1103/PhysRevLett.101.080402}.

\bibitem{Joglekar2013}
Y.~N. Joglekar, C.~Thompson, D.~D. Scott, G.~Vemuri.
\newblock Optical waveguide arrays: quantum effects and pt symmetry breaking.
\newblock \textit{The European Physical Journal Applied Physics}
  \textbf{63}(3):30001, 2013.
\newblock \bibdoi{10.1051/epjap/2013130240}.

\bibitem{Kato1995}
T.~Kato.
\newblock \textit{Perturbation Theory for Linear Operators}.
\newblock Springer Berlin Heidelberg, 1995.
\newblock \bibdoi{10.1007/978-3-642-66282-9}.

\bibitem{Ruter2010}
C.~E. R{\"{u}}ter, K.~G. Makris, R.~El-Ganainy, et~al.
\newblock {Observation of parity-time symmetry in optics}.
\newblock \textit{Nature Physics} \textbf{6}(3):192--195, 2010.
\newblock \eprint{1003.4968} \bibdoi{10.1038/nphys1515}.

\bibitem{Regensburger2012}
A.~Regensburger, C.~Bersch, M.-A. Miri, et~al.
\newblock Parity{\textendash}time synthetic photonic lattices.
\newblock \textit{Nature} \textbf{488}(7410):167--171, 2012.
\newblock \bibdoi{10.1038/nature11298}.

\bibitem{Hodaei2014}
H.~Hodaei, M.-A. Miri, M.~Heinrich, et~al.
\newblock Parity-time-symmetric microring lasers.
\newblock \textit{Science} \textbf{346}(6212):975--978, 2014.
\newblock \bibdoi{10.1126/science.1258480}.

\bibitem{Peng2014}
B.~Peng, {\c{S}}.~K. \"{O}zdemir, F.~Lei, et~al.
\newblock Parity{\textendash}time-symmetric whispering-gallery microcavities.
\newblock \textit{Nature Physics} \textbf{10}(5):394--398, 2014.
\newblock \bibdoi{10.1038/nphys2927}.

\bibitem{Chang2014}
L.~Chang, X.~Jiang, S.~Hua, et~al.
\newblock Parity{\textendash}time symmetry and variable optical isolation in
  active{\textendash}passive-coupled microresonators.
\newblock \textit{Nature Photonics} \textbf{8}(7):524--529, 2014.
\newblock \bibdoi{10.1038/nphoton.2014.133}.

\bibitem{Schindler2011}
J.~Schindler, A.~Li, M.~C. Zheng, et~al.
\newblock {Experimental study of active LRC circuits with PT symmetries}.
\newblock \textit{Physical Review A - Atomic, Molecular, and Optical Physics}
  \textbf{84}(4):1--5, 2011.
\newblock \bibdoi{10.1103/PhysRevA.84.040101}.

\bibitem{Wang2020}
T.~Wang, J.~Fang, Z.~Xie, et~al.
\newblock Observation of two pt transitions in an electric circuit with
  balanced gain and loss.
\newblock \textit{The European Physical Journal D} \textbf{74}(8), 2020.
\newblock \bibdoi{10.1140/epjd/e2020-10131-7}.

\bibitem{Quirozjuarez2021}
M.~A. Quiroz-Juárez, K.~S. Agarwal, Z.~A. Cochran, et~al.
\newblock On-demand parity-time symmetry in a lone oscillator through complex,
  synthetic gauge fields, 2021.
\newblock \eprint{2109.03846}.

\bibitem{Bender2013}
C.~M. Bender, B.~K. Berntson, D.~Parker, E.~Samuel.
\newblock Observation of pt phase transition in a simple mechanical system.
\newblock \textit{American Journal of Physics} \textbf{81}(3):173--179, 2013.
\newblock \bibdoi{10.1119/1.4789549}.

\bibitem{Duchesne2009}
D.~Duchesne, V.~Aimez, R.~Morandotti, et~al.
\newblock {Observation of PT -Symmetry Breaking in Complex Optical Potentials}.
\newblock \textit{Physical Review Letters} \textbf{103}(9):1--4, 2009.
\newblock \bibdoi{10.1103/physrevlett.103.093902}.

\bibitem{LeonMontiel2018}
R.~de~J.~Le{\'{o}}n-Montiel, M.~A. Quiroz-Ju{\'{a}}rez, J.~L.
  Dom{\'{\i}}nguez-Ju{\'{a}}rez, et~al.
\newblock Observation of slowly decaying eigenmodes without exceptional points
  in floquet dissipative synthetic circuits.
\newblock \textit{Communications Physics} \textbf{1}(1), 2018.
\newblock \bibdoi{10.1038/s42005-018-0087-3}.

\bibitem{Joglekar2018}
Y.~N. Joglekar, A.~K. Harter.
\newblock Passive parity-time-symmetry-breaking transitions without exceptional
  points in dissipative photonic systems [invited].
\newblock \textit{Photonics Research} \textbf{6}(8):A51, 2018.
\newblock \bibdoi{10.1364/prj.6.000a51}.

\bibitem{Wu2019}
Y.~Wu, W.~Liu, J.~Geng, et~al.
\newblock Observation of parity-time symmetry breaking in a single-spin system.
\newblock \textit{Science} \textbf{364}(6443):878--880, 2019.
\newblock \bibdoi{10.1126/science.aaw8205}.

\bibitem{naghiloo2019}
M.~Naghiloo, M.~Abbasi, Y.~N. Joglekar, K.~W. Murch.
\newblock Quantum state tomography across the exceptional point in a single
  dissipative qubit.
\newblock \textit{Nature Physics} \textbf{15}(12):1232--1236, 2019.
\newblock \bibdoi{10.1038/s41567-019-0652-z}.

\bibitem{Li2019}
J.~Li, A.~K. Harter, J.~Liu, et~al.
\newblock Observation of parity-time symmetry breaking transitions in a
  dissipative floquet system of ultracold atoms.
\newblock \textit{Nature Communications} \textbf{10}(1), 2019.
\newblock \bibdoi{10.1038/s41467-019-08596-1}.

\bibitem{klauck2019}
F.~Klauck, L.~Teuber, M.~Ornigotti, et~al.
\newblock Observation of $\mathcal{PT}$-symmetric quantum interference.
\newblock \textit{Nature Photonics} \textbf{13}(12):883--887, 2019.
\newblock \bibdoi{10.1038/s41566-019-0517-0}.

\bibitem{Ruzicka2021}
F.~Ruzicka, K.~S. Agarwal, Y.~N. Joglekar.
\newblock Conserved quantities, exceptional points, and antilinear symmetries
  in non-hermitian systems.
\newblock \textit{Journal of Physics: Conference Series}
  \textbf{2038}(1):012021, 2021.
\newblock \bibdoi{10.1088/1742-6596/2038/1/012021}.

\bibitem{Bian2020}
Z.~Bian, L.~Xiao, K.~Wang, et~al.
\newblock Conserved quantities in parity-time symmetric systems.
\newblock \textit{Physical Review Research} \textbf{2}(2), 2020.
\newblock \bibdoi{10.1103/physrevresearch.2.022039}.

\bibitem{Berry2008}
M.~V. Berry.
\newblock Optical lattices with {PT} symmetry are not transparent.
\newblock \textit{Journal of Physics A: Mathematical and Theoretical}
  \textbf{41}(24):244007, 2008.
\newblock \bibdoi{10.1088/1751-8113/41/24/244007}.

\bibitem{Teimourpour2014}
M.~H. Teimourpour, R.~El-Ganainy, A.~Eisfeld, et~al.
\newblock Light transport in $\mathcal{PT}$-invariant photonic structures with
  hidden symmetries.
\newblock \textit{Phys Rev A} \textbf{90}:053817, 2014.
\newblock \bibdoi{10.1103/PhysRevA.90.053817}.

\bibitem{Gorini1976}
V.~Gorini, A.~Kossakowski, E.~C.~G. Sudarshan.
\newblock Completely positive dynamical semigroups of n-level systems.
\newblock \textit{Journal of Mathematical Physics} \textbf{17}(5):821, 1976.
\newblock \bibdoi{10.1063/1.522979}.

\bibitem{Lindblad1976}
G.~Lindblad.
\newblock On the generators of quantum dynamical semigroups.
\newblock \textit{Comm Math Phys} \textbf{48}(2):119--130, 1976.

\bibitem{Ban1993}
M.~Ban.
\newblock Lie-algebra methods in quantum optics: The liouville-space
  formulation.
\newblock \textit{Physical Review A} \textbf{47}(6):5093--5119, 1993.
\newblock \bibdoi{10.1103/physreva.47.5093}.

\bibitem{Albert2014}
V.~V. Albert, L.~Jiang.
\newblock Symmetries and conserved quantities in lindblad master equations.
\newblock \textit{Phys Rev A} \textbf{89}:022118, 2014.
\newblock \bibdoi{10.1103/PhysRevA.89.022118}.

\bibitem{Manzano2020}
D.~Manzano.
\newblock A short introduction to the lindblad master equation.
\newblock \textit{{AIP} Advances} \textbf{10}(2):025106, 2020.
\newblock \bibdoi{10.1063/1.5115323}.

\bibitem{Gunderson2021}
J.~Gunderson, J.~Muldoon, K.~W. Murch, Y.~N. Joglekar.
\newblock Floquet exceptional contours in lindblad dynamics with time-periodic
  drive and dissipation.
\newblock \textit{Phys Rev A} \textbf{103}:023718, 2021.
\newblock \bibdoi{10.1103/PhysRevA.103.023718}.

\bibitem{Joglekar2014}
Y.~N. Joglekar, R.~Marathe, P.~Durganandini, R.~K. Pathak.
\newblock {PTspectroscopy} of the rabi problem.
\newblock \textit{Physical Review A} \textbf{90}(4):040101, 2014.
\newblock \bibdoi{10.1103/physreva.90.040101}.

\bibitem{Lee2015}
T.~E. Lee, Y.~N. Joglekar.
\newblock $\mathcal{PT}$-symmetric rabi model: Perturbation theory.
\newblock \textit{Phys Rev A} \textbf{92}:042103, 2015.
\newblock \bibdoi{10.1103/PhysRevA.92.042103}.

\bibitem{Hanggi1998}
P.~Hanggi.
\newblock \textit{Diven quantum systems}, 1998 (accessed October 31, 2020).

\bibitem{Harter2020}
A.~K. Harter, Y.~N. Joglekar.
\newblock Connecting active and passive
  {\textdollar}{\textbackslash}mathcal$\lbrace${PT}$\rbrace${\textdollar}-symmetric
  floquet modulation models.
\newblock \textit{Progress of Theoretical and Experimental Physics}
  \textbf{2020}(12), 2020.
\newblock \bibdoi{10.1093/ptep/ptaa181}.

\bibitem{Peng2016}
P.~Peng, W.~Cao, C.~Shen, et~al.
\newblock Anti-parity{\textendash}time symmetry with flying atoms.
\newblock \textit{Nature Physics} \textbf{12}(12):1139--1145, 2016.
\newblock \bibdoi{10.1038/nphys3842}.

\bibitem{Choi2018}
Y.~Choi, C.~Hahn, J.~W. Yoon, S.~H. Song.
\newblock Observation of an anti-pt-symmetric exceptional point and
  energy-difference conserving dynamics in electrical circuit resonators.
\newblock \textit{Nature Communications} \textbf{9}(1), 2018.
\newblock \bibdoi{10.1038/s41467-018-04690-y}.

\bibitem{Zhang2020}
F.~Zhang, Y.~Feng, X.~Chen, et~al.
\newblock Synthetic anti-pt symmetry in a single microcavity.
\newblock \textit{Phys Rev Lett} \textbf{124}:053901, 2020.
\newblock \bibdoi{10.1103/PhysRevLett.124.053901}.

\bibitem{Longhi2019}
S.~Longhi, E.~Pinotti.
\newblock Anyonic $\mathcal{PT}$ symmetry, drifting potentials and
  non-hermitian delocalization.
\newblock \textit{{EPL} (Europhysics Letters)} \textbf{125}(1):10006, 2019.
\newblock \bibdoi{10.1209/0295-5075/125/10006}.

\bibitem{Arwas2021}
G.~Arwas, S.~Gadasi, I.~Gershenzon, et~al.
\newblock Anyonic parity-time symmetric laser, 2021.
\newblock \eprint{2103.15359}.

\end{thebibliography}

\end{document}